%% file: main.tex
\documentclass[%
 aip,
 amsmath,amssymb,
 reprint,%
]{revtex4-1}

\input{tex/auxil}

\begin{document}

\title{Observation of unconventional spin polarization induced spin orbit torque in L1$_2$-ordered antiferromagnetic \mnpt{} thin films
}
\input{tex/00_authors}
\input{tex/00_abstract}
\date{\today}

\maketitle
\input{tex/01_intro}


\input{tex/03_conclusion}

\input{tex/10_ackn}
\bibliography{refs}

\end{document}

%% file: tex/auxil.tex
\usepackage{graphicx}
\graphicspath{{./figs/}}

\usepackage{dcolumn}
\usepackage{bm}
\usepackage{hyperref}
\usepackage[svgnames]{xcolor}
\usepackage[mathlines]{lineno}
\usepackage{miller}

\usepackage{siunitx}
\DeclareSIUnit{\dbm}{dBm}

\newcommand{\mnpt}{M\lowercase{n}$_3$P\lowercase{t}}

%% file: tex/00_authors.tex
\author{Longjie Yu}
\email{yu.longjie.t7dc.tohoku.ac.jp}
\affiliation{Department of Applied Physics, Tohoku University, Sendai 980-8579, Japan}

\author{Shutaro Karube}
\affiliation{Department of Materials Science, Tohoku University, Sendai 980-8579, Japan}
\affiliation{Center for Spintronics Research Network, Sendai 980-8579, Japan}

\author{Min Liu}
\affiliation{Department of Applied Physics, Tohoku University, Sendai 980-8579, Japan}

\author{Masakiyo Tsunoda}
\affiliation{Center for Spintronics Research Network, Sendai 980-8579, Japan}
\affiliation{Department of Electronic Engineering, Tohoku University, Sendai 980-8579, Japan}

\author{Mikihiko Oogane}
\affiliation{Department of Applied Physics, Tohoku University, Sendai 980-8579, Japan}
\affiliation{Center for Spintronics Research Network, Sendai 980-8579, Japan}
\affiliation{Center for Science and Innovation in Spintronics, Sendai 980-8579, Japan}

\author{Yasuo Ando}
\affiliation{Department of Applied Physics, Tohoku University, Sendai 980-8579, Japan}
\affiliation{Center for Spintronics Research Network, Sendai 980-8579, Japan}
\affiliation{Center for Science and Innovation in Spintronics, Sendai 980-8579, Japan}

%% file: tex/00_abstract.tex
\begin{abstract}
    Non-collinear antiferromagnets exhibits richer magneto-transport properties due to the topologically nontrivial spin structure they possess compared to conventional nonmagnetic materials, which allows us to manipulate the charge-spin conversion more freely by taking advantage of the chirality. In this work, we explore the unconventional spin orbit torque of L1$_2$-ordered \mnpt{} with a triangular spin structure. We observed an unconventional spin orbit torque along the $\mathbf{x}$-direction for the (001)-oriented L1$_2$ \mnpt{}, and found that it has a unique sign reversal behavior relative to the crystalline orientation. This generation of unconventional spin orbit torque for L1$_2$-ordered \mnpt{} can be interpreted as stemming from the magnetic spin Hall effect. This report help clarify the correlation between the topologically nontrivial spin structure and charge-spin conversion in non-collinear antiferromagnets.
\end{abstract}

%% file: tex/01_intro.tex
Many newly discovered phenomena associated with the topology of magnetism make antiferromagnets (AFMs) with topologically nontrivial spin structures a rich playground for the investigation of unique topological behaviors, as well as promising candidates for energy-efficient microelectronic applications~\cite{baltz2018antiferromagnetic, jungfleisch2018perspectives, vsmejkal2018topological}. The anomalous transport properties of electrons and magnons that are affected by spin structure induced non-vanishing Berry curvatures~\cite{chen2014anomalous, nakatsuji2015large} or chirality-related Dzyaloshinskii-Moriya interaction~\cite{fert2017magnetic}, known as the anomalous Hall effect (AHE) and magnetic skyrmions respectively, make it possible to control the charge or magnon transport in AFMs efficiently and stably. AFMs have also attracted interest because it is possible to not only use them as efficient spin current sources~\cite{zhang2014spin,zhang2016giant,zhou2019large} but also to modulate their spin structures electrically~\cite{wadley2016electrical, chen2019electric, takeuchi2021chiral}, which leads to great application potential. More recently, the discovery of magnetic spin Hall effect (MSHE)~\cite{kimata2019magnetic,holanda2020magnetic, mook2020origin} in non-collinear AFMs with topological triangular spin structures and the observation of Berry curvature-induced spin-orbit torque (SOT)~\cite{chen2021observation, kurebayashi2014antidamping} in collinear AFM with a parallel spin structure have been highlighted, both of which provide the opportunity to manipulate the spin current and generate spin polarization by tuning the spin structures of AFMs.

Although the charge-to-spin conversion phenomenon of MSHE is similar to that of the intrinsic conventional spin Hall effect (SHE), the time-reversal symmetry of MSHE is odd and of extrinsic origin, which makes it a reactive counterpart of SHE and demonstrative of a dissipative nature~\cite{vzelezny2017spin, kimata2019magnetic, holanda2020magnetic, mook2020origin}. The odd time-reversal symmetry can be regarded as the result of magnetic-order parameter reversal, which indicates that the spin polarization $\mathbf{\sigma}$ generated by MSHE is dictated by the magnetic order, thereby allowing an in-plane component in the form of scattering plane created by the charge current and spin current. In other words, in current-induced SOT, MSHE is expected to generate an additional NO-$\mathbf{y}$ spin polarization, \emph{i.e.}, $\mathbf{x}$-polarization and $\mathbf{z}$-polarization when the charge current is flowing along the in-plane direction ($\mathbf{x}$ direction), whereas only $\mathbf{y}$-polarization is allowed to exist for SHE due to the restriction of symmetry~\cite{kimata2019magnetic, nan2020controlling, chen2021observation}. Analogous to the $\mathbf{y}$-polarization which generated by SHE will give rise to a conventional damping-like (DL) torque in an adjacent ferromagnet (FM) of the form $\mathbf{m}\times(\mathbf{m}\times \mathbf{y})$, where $\mathbf{m}$ denotes the magnetization vector, it is expected that the unconventional DL torque of $\mathbf{x}$-polarization and $\mathbf{z}$-polarization generated by MSHE will take the forms of  $\mathbf{m}\times(\mathbf{m}\times \mathbf{x})$ and $\mathbf{m}\times(\mathbf{m}\times \mathbf{z})$, respectively. This means that MSHE-induced SOT can be investigated quantitatively under the condition of ferromagnetic resonance. In doing so, we hope to contribute to a deeper understanding of unconventional SOT induced by MSHE~\cite{macneill2017control}.

\begin{figure}
	\includegraphics[width=0.5\textwidth]{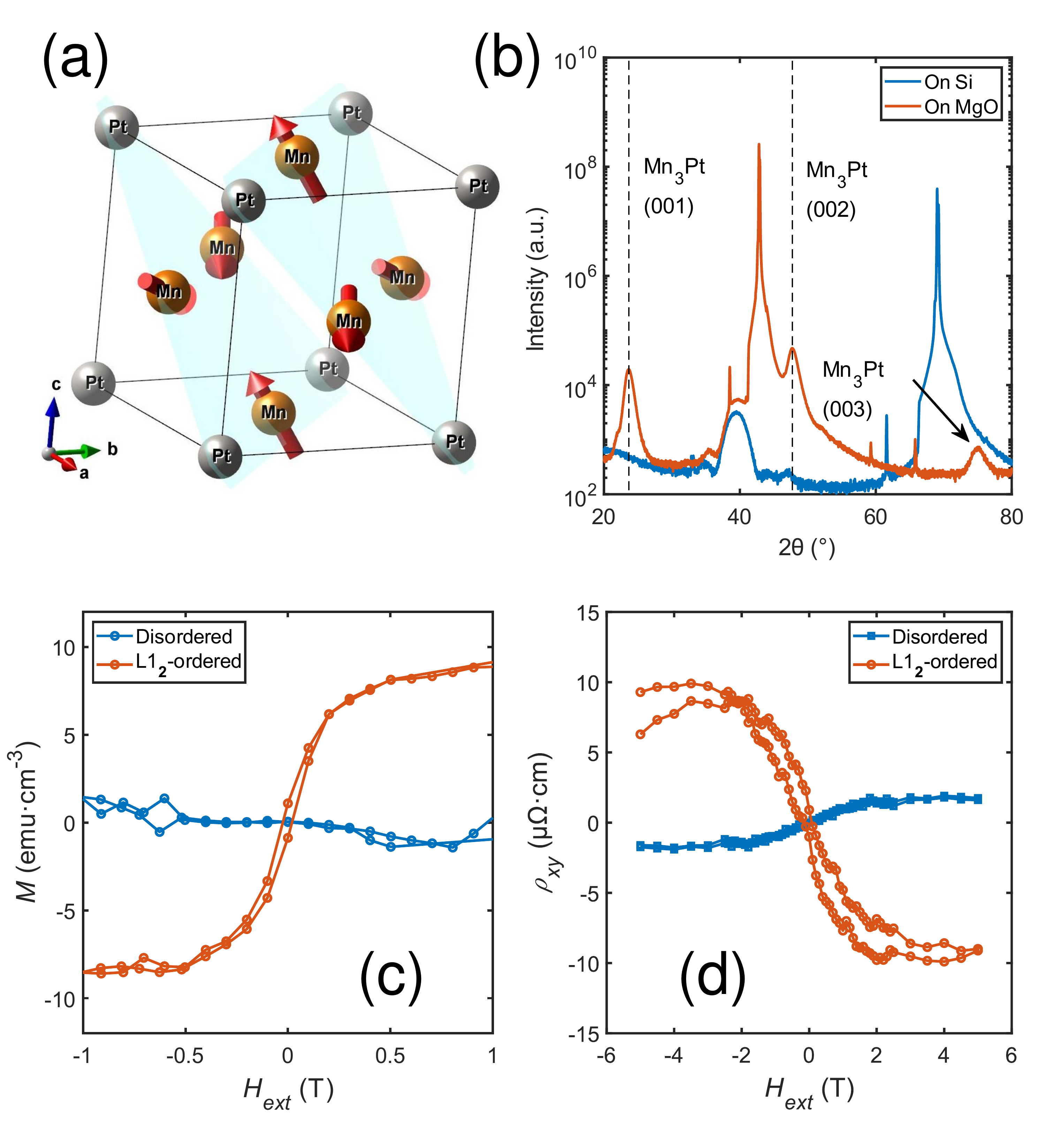}
	\caption{(a) Crystal and magnetic structure of L1$_2$-ordered \mnpt{}. Note that the Mn atoms in the unit cell of \mnpt{} construct a kagome lattice in the (111) plane, where the spin structure is triangular and the direction of spins is either all-in or all-out. (b) X-ray diffraction patterns measured for the 15 \si{\nm}-thick \mnpt{} deposited on MgO and Si substrates. (c) Measured magnetic hysteresis curves for the \mnpt{} with and without L1$_2$-ordered structure. A weak magnetization can only be detected for L1$_2$-ordered \mnpt{}. (d) AHE of samples in (c).}
	\label{fig:1}
\end{figure}
\begin{figure*}[t]
        \includegraphics[width=1.0\textwidth]{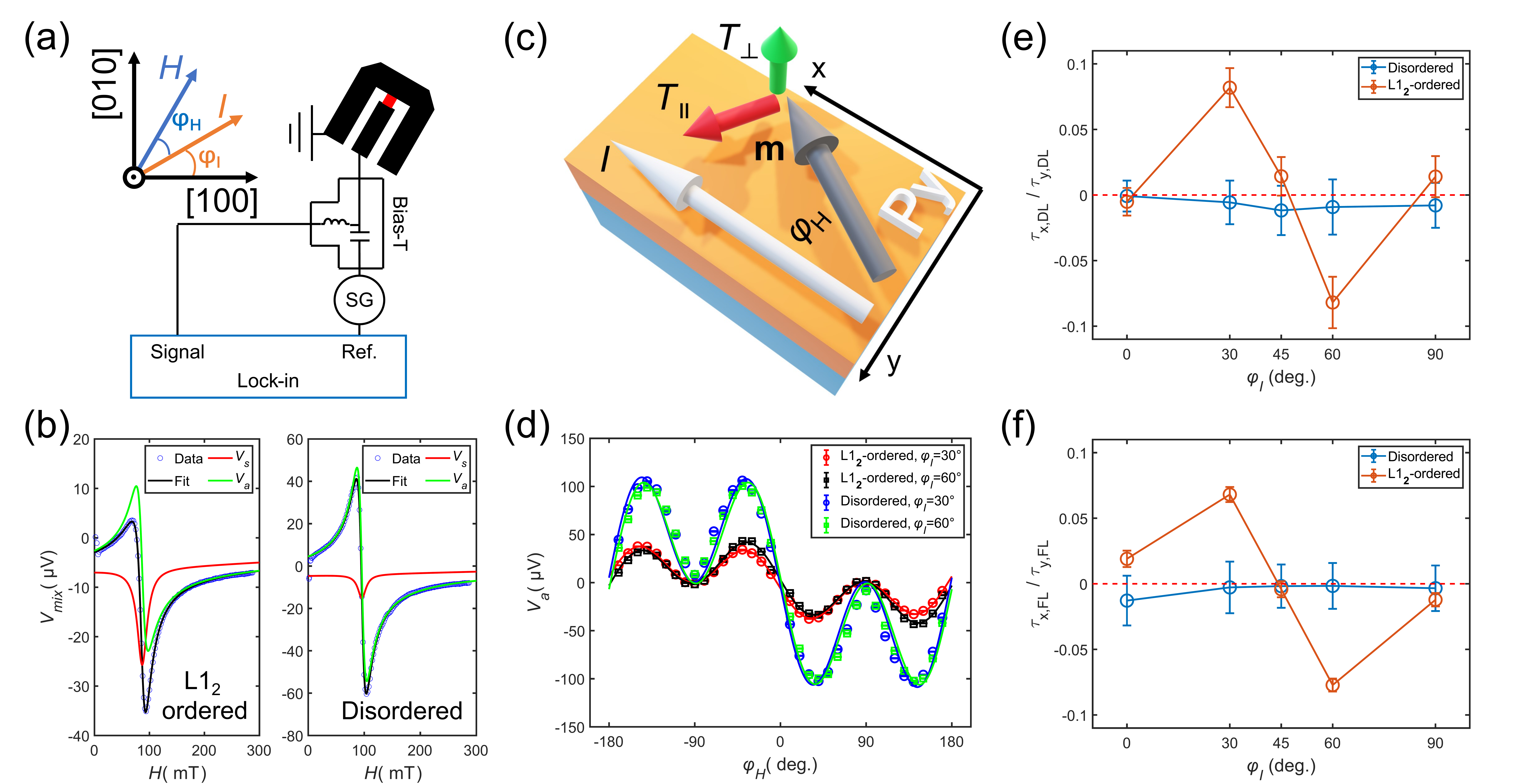}
        \caption{(a) Schematic illustration of ST-FMR measurement. Coordinate is defined with respect to the lattice of \mnpt{}. The red strip represents \mnpt{} / Py bilayer. (b) Measured $V_{mix}$ for L1$_2$-ordered \mnpt{} / Py and disordered \mnpt{} / Py at 9 \si{\GHz} with $\phi_{H}=40\si{\degree}$. The applied microwave power is 6 \si{\dbm}. (c) Schematic of the sample geometry and SOTs acting on Py: $\mathbf{\tau_{\parallel}}$ is composed of $\tau_{x,DL}$ and $\tau_{y,DL}$, while $\mathbf{\tau_{\perp}}$ is composed of $\tau_{x,FL}$ and $\tau_{y,FL}$. (d) Antisymmetric ST-FMR component ($V_a$) of L1$_2$-ordered \mnpt{} / Py and disordered \mnpt{} / Py with different $\phi_{I}$ as a funtion of $\phi_{H}$. The error bars represent fitting standard deviations. (e, f) Torque ratios $\tau_{x,DL} / \tau_{y,DL}$ and $\tau_{x,FL} / \tau_{y,FL}$ for L1$_2$-ordered \mnpt{} / Py and disordered \mnpt{} / Py as a function of $\phi_{I}$.}
        \label{fig:2}
\end{figure*}

The non-collinear AFM L1$_2$ \mnpt{} that we focus on this paper is an antiferromagnetic material with a topologically nontrivial spin structure~\cite{kren1968magnetic, chen2014anomalous}. The crystal structure of \mnpt{} is L1$_2$ (space group: $Pm\overline{3}m$), and the N\'eel temperature is around 475\si{\kelvin}. At room temperature, \mnpt{} has a 120\si{\degree} triangular $\Gamma_{4g}$ spin structure lying in a (111) kagome plane where the spin configuration is all-in or all-out, which is similar to Mn$_3$Ir (see Fig.~\ref{fig:1}(a))~\cite{kohn2013antiferromagnetic}. \mnpt{} has so far been experimentally demonstrated to exhibit a relatively large AHE~\cite{liu2018electrical} and spin Hall angle~\cite{zhang2014spin}. In addition, an intermetallic compound with the same $\Gamma_{4g}$ spin structure also shows facet-dependence of spin Hall conductivity~\cite{zhang2016giant} and out-of-plane ($\mathbf{z}$ direction) spin accumulation~\cite{liu2019current}. In this work, we use a spin-torque ferromagnetic resonance (ST-FMR) technique to explore the unconventional SOT in an L1$_2$-ordered epitaxial \mnpt{} / Ni$_{80}$Fe$_{20}$ (Py) heterostructure. We observed the NO-$\mathbf{y}$ spin-polarization-induced SOT in L1$_2$-ordered \mnpt{}, which is absent in poly-\mnpt{} whose lattice structure as well as spin structure are regarded as disordered. Such an SOT can be associated with the contribution from the MSHE and the configuration of AFM domains.

\mnpt{} thin films with the thickness of 15\si{\nm} were grown by magnetron sputtering on a MgO (001) and a thermally oxidized Si substrate with the substrate temperature $T_s$ of 450\si{\degreeCelsius} and RT, respectively. Subsequently, a thin layer of  MgAl$_2$O$_4$ (2.5 \si{\nm}) was deposited at RT to prevent oxidation. The out-of-plane X-ray diffraction patterns for \mnpt{} fabricated on the MgO and Si substrates are shown in Fig.~\ref{fig:1}(b). (001) and (003) superlattice peaks originating from the L1$_2$ structure could be observed clearly only for the \mnpt{} deposited on MgO, while they disappeared for the one deposited on Si. Such a result indicates that only \mnpt{} deposited on MgO forms a well-ordered L1$_2$ structure, while the lattice structure of \mnpt{} deposited on Si is disordered. Therefore, we set \mnpt{} deposited on Si substrate as a control sample. In addition, we also note that the aforementioned $\Gamma_{4g}$ spin structure only exists in the L1$_2$ structure. 

Next, to determine whether L1$_2$ \mnpt{} had a triangular spin structure, we characterized the magnetic and magneto-transport properties with magnetic hysteresis curves and performed Hall measurement for the \mnpt{} single film in an out-of-plane applied field. To examine the magnetic properties, we compared the hysteresis curves of L1$_2$-ordered \mnpt{} and disordered \mnpt{} at 300\si{\kelvin}, as shown in Fig.~\ref{fig:1}(c). A weak but nonzero net magnetization $M$ (about 9 emu·cm$^{-3}$)  was detected along the (001) direction of the L1$_2$ \mnpt{} film due to triangle spin canting, which is consistent with previous theoretical~\cite{chen2014anomalous} and experimental works~\cite{liu2018electrical}. The magnitude of $M$ we measured is also similar to that for L1$_2$-ordered Mn$_3$Ir~\cite{iwaki2020large}. In contrast, such a small magnetization could not be observed for the disordered control sample. For transport properties, we patterned the films by photo-lithography and ion-beam etching into a Hall bar with the channel width of 10 \si{\um} and measured transverse resistivity $\rho_{xy}$ by applying current along [100] direction. Figure~\ref{fig:1}(d) shows $\rho_{xy}$ as a function of the external field at 300\si{\kelvin}. Negative Hall-resistance loops were observed for L1$_2$-ordered \mnpt{}, which means the signature of AHE exists in L1$_2$-ordered \mnpt{} and corresponds to the previous observation in the Mn$_3$X (X = Sn, Ge, Ga, Ir, Pt) family of AFM materials~\cite{ikeda2018anomalous, iwaki2020large, kiyohara2016giant, liu2018electrical, mukherjee2021sign}. 

We next evaluated the symmetry of SOT and examined the SOT components quantitatively using an ST-FMR technique~\cite{macneill2017control, nan2020controlling, chen2021observation, liu2011spin}. The stack structures we measured were Si or MgO / \mnpt{} (15 \si{\nm}) / Py (8 \si{\nm}) / MgAl$_2$O$_4$ (2.5 \si{\nm}). Figure~\ref{fig:2}(a) shows the schematic geometry of the ST-FMR measurement. A microwave current $I$ was applied to \mnpt{}, torques were generated therefore exerted to the Py. An in-plane external field was simultaneously swept at an angle $\phi_{H}$ relative to $I$, after which a mix voltage $V_{mix}$ modulated by the anisotropic magnetoresistance (AMR) of Py could be detected by a lock-in amplifier. Such a $V_{mix}$ can be fitted as $V_{mix}=V_sF_s+V_aF_a$, where $F_s$ and $F_a$ denote the symmetric and antisymmetric Lorentzian functions, and $V_s$ and $V_a$ denote the amplitudes of $F_s$ and $F_a$, respectively. Figure~\ref{fig:2}(b) shows the ST-FMR spectra of L1$_2$-ordered \mnpt{} / Py and disordered \mnpt{} / Py at 9 \si{\GHz} with $\phi_{H}=40\si{\degree}$ and symmetric and antisymmetric components separated by fitting to $V_{mix}=V_sF_s+V_aF_a$. The data points were fitted well by the equation above. In order to evaluate the possible DL torque and FL torque generated by NO-$\mathbf{y}$ polarization due to MSHE, we also performed angle-dependent ST-FMR measurements with respect to $\phi_{H}$, which a reliable approach for identifying unconventional torques. According to the symmetric analysis of spin polarization-induced SOTs that exert to the Py layer, the in-plane torque $\mathbf{\tau_{\parallel}}$ and out-of-plane torque $\mathbf{\tau_{\perp}}$ components of SOT are proportional to $V_s$ and $V_a$, which can be expressed as~\cite{macneill2017control, nan2020controlling} 
\begin{equation}\label{eq:1}
\begin{split}
    V_s ( \phi_H  )&\propto\sin  ( 2\phi_H  )\mathbf{\tau_{\parallel}}\\
    &=\sin ( 2\phi_H   ) [ \tau_{x,DL}\sin (\phi_H   ) +\tau_{y,DL}\cos (\phi_H   )\\
    &+\tau_{z,FL} ],
\end{split}
\end{equation}
\begin{equation}\label{eq:2}
\begin{split}
    V_a ( \phi_H  )&\propto\sin  ( 2\phi_H  )\mathbf{\tau_{\perp}}\\
    &=\sin ( 2\phi_H   ) [ \tau_{x,FL}\sin (\phi_H   ) +\tau_{y,FL}\cos (\phi_H   )\\
    &+\tau_{z,DL} ],
\end{split}
\end{equation}
where $\tau_{x,DL}$, $\tau_{x,FL}$, $\tau_{y,DL}$, $\tau_{y,FL}$, $\tau_{z,DL}$, $\tau_{z,FL}$ are the DL / FL torques generated by spin currents that are polarized along the $\mathbf{x}$, $\mathbf{y}$, $\mathbf{z}$ directions, respectively. Note that effects such as Rashba-like and Dresselhaus-like fields and exchange coupling at the FM and AFM interface may contribute to the above torques, as discussed below. Figure.~\ref{fig:2}(c) depicts the components of SOTs exerting on Py. 

\begin{figure*}[ht]
        \includegraphics[width=1.0\textwidth]{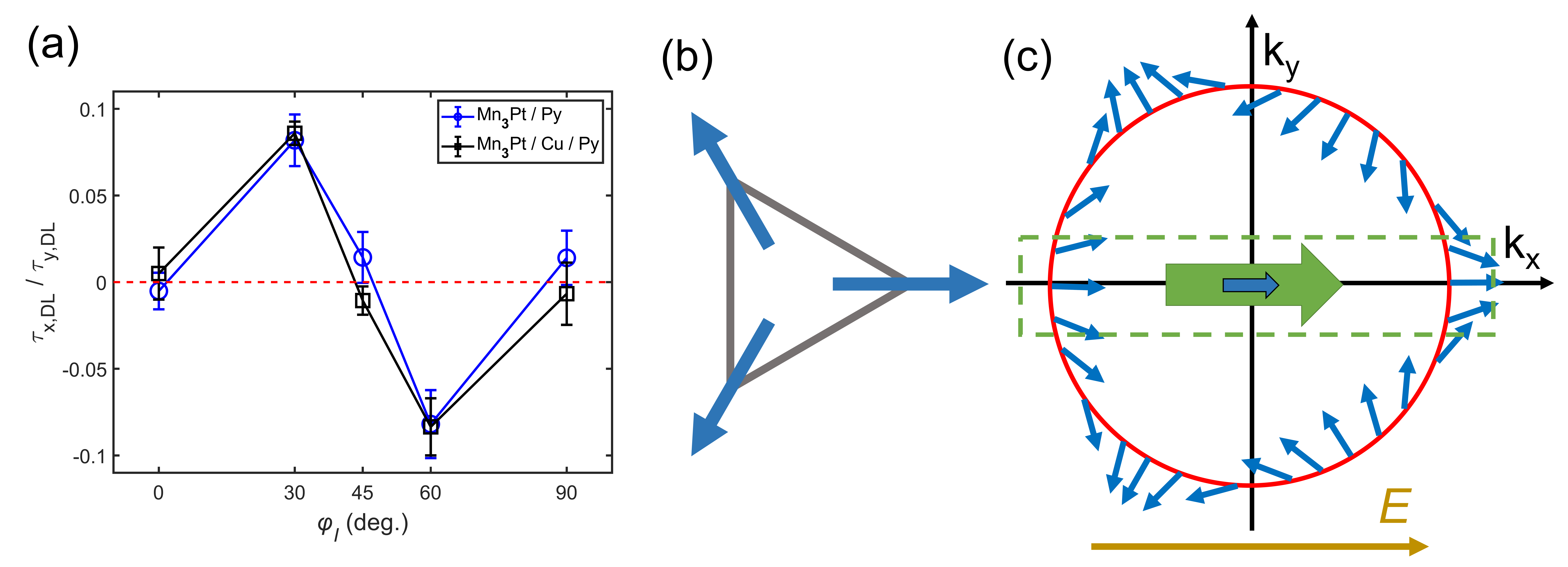}
        \caption{(a) Comparison of $\tau_{x,DL} / \tau_{y,DL}$ dependence of $\phi_{I}$ for L1$_2$-ordered \mnpt{} with or without Cu interlayer. (b) $\Gamma_{4g}$ magnetic configuration of the L1$_2$-ordered \mnpt{} possesses, and (c) its corresponding spin structure in momentum space. In the picture of MSHE~\cite{vzelezny2017spin}, a longitudinal spin current will be generated as a result of the non-collinear spin structure breaking the spin conservation.}
        \label{fig:3}
\end{figure*}

We further investigated the unconventional torques ($\tau_{x}$ and $\tau_{z}$) by patterning a series of microstrips on the same sample with different $\phi_{I}$, where $\phi_{I}$ is defined as the angle between [100] orientation and $I$, thereby measuring the $\phi_{H}$ dependence of $V_a$ and $V_s$. In Fig.~\ref{fig:2}(d), we compare the $V_a$ dependence of $\phi_{H}$ for L1$_2$-ordered \mnpt{} and disordered \mnpt{} with different $\phi_{I}$. Two phenomena could be observed clearly for \mnpt{} deposited on MgO, which is L1$_2$-ordered: (1) there were $\sin ( 2\phi_H   )$ components induced by $\tau_{x,FL}$ in the two microstrips, giving rise to the broken centrosymmetry of $V_a$, and (2) on the one side of plus $\phi_{H}$ or minus $\phi_{H}$, the relative magnitudes of two peaks looks difficult, and the symmetry were reversed from $\phi_{I}=30\si{\degree}$ to $\phi_{I}=60\si{\degree}$ due to the sign reversal of $\tau_{x,FL}$. For instance, when $\phi_{H}$ decreasing from -180\si{\degree} to 0\si{\degree}, peak values decreased from 37.4 \si{\uA} to 33.3 \si{\uA} for $\phi_{I}=30\si{\degree}$, while peak values increased from 32.7 \si{\uA} to 41.9 \si{\uA} for $\phi_{I}=60\si{\degree}$. In contrast, these two phenomena could not be found for disordered \mnpt{}. It is also worthwhile to note that although the magnitude of the $V_a$ of L1$_2$-ordered \mnpt{} and disordered \mnpt{} are obviously different due to the distinct impedance mismatch, it makes no difference to the symmetry, as discussed above. $V_s$ also exhibited a similar behavior attribute to the existence of the $\sin ( 2\phi_H  )$ contribution induced by $\tau_{x,DL}$ although not shown here. Such a result demonstrates a strong correlation between unconventional torques and the triangular spin structure. We therefore measured the $V_a$ and $V_s$ as a function of $\phi_{H}$ from $\phi_{I}=0\si{\degree}$ to $\phi_{I}=90\si{\degree}$and fitted them to Eq.(\ref{eq:1}, \ref{eq:2}), thereby extracting the associated fitting parameters $\tau_{x,DL}$, $\tau_{x,FL}$, $\tau_{y,FL}$ and plotting the ratios of  $\tau_{x,DL} / \tau_{y,DL}$ and $\tau_{x,FL} / \tau_{y,FL}$ as a function of $\phi_{I}$, as shown in Fig.~\ref{fig:2}(e) and (f). Such a torque ratio can help us to quantitatively estimate the $\mathbf{x}$ component and $\mathbf{y}$ component that contribute to $V_s$ and $V_a$. The FL torque generated by $\mathbf{y}$-polarization is negligible so that Oersted field contributes $\tau_{y,FL}$ mainly. Detectable $\tau_{x,DL} / \tau_{y,DL}$ of about 0.08 and $\tau_{x,FL} / \tau_{y,FL}$ of about 0.07 were observed for L1$_2$-ordered \mnpt{}, and a sign inversion behavior was clearly exhibited for both ratios, which may have originated from the generation and polarization reversal of spin polarization along the $\mathbf{x}$ direction (current direction). However, both the detectable $\tau_{x}$ and the sign inversion behavior vanish for the disordered one. Notably, compared to the conventional $\tau_{y}$, which yields a Rashba-like field with the same Rashiba-like symmetry, the unconventional FL and DL torques $\tau_{x,FL}$ and $\tau_{x,DL}$ that have been observed above correspond to a Dresselhaus-like symmetry~\cite{kurebayashi2014antidamping, ciccarelli2016room}.

The same angular-dependent experiment was carried out in MgO / \mnpt{} (15 \si{\nm}) / Cu (2 \si{\nm}) / Py (8 \si{\nm}) stacks with a Cu insertion layer to rule out possible effects caused by interlayer exchange coupling. We compared torque ratios $\tau_{x,DL} / \tau_{y,DL}$ for sample with and without a Cu layer, as shown in Fig.~\ref{fig:3}(a). Remarkably, the variation as well as the values of $\tau_{x,DL} / \tau_{y,DL}$ for the samples with a Cu spacer were quite similar to its counterpart without Cu, indicating that the interlayer effects between L1$_2$-ordered \mnpt{} and Py are not likely to account for the appearance of $\mathbf{x}$-polarization. The appearance of $\mathbf{x}$-polarization in our \mnpt{} sample with a triangular $\Gamma_{4g}$ spin structure can be interpreted phenomenologically by the fact that spin-momentum locking yields MSHE~\cite{vzelezny2017spin, mook2020origin}. Recent reports have demonstrated that the momentum space of non-collinear AFM is related strongly to its spin structure. Figure.~\ref{fig:3}(b, c) show the schematics of the triangular $\Gamma_{4g}$ magnetic configuration in real space and the corresponding spin structure in momentum space, respectively. Upon applying an electric field, the distribution of the electrons at the Fermi level will changed, and a spin current with $\mathbf{x}$-polarization along the $\mathbf{x}$ direction will be generated for both states of $k_y>0$ and $k_y<0$. The property of the canted kagome plane in our (001)-oriented L1$_2$ \mnpt{} sample will give rise to the detection of $\mathbf{x}$-polarization-induced $\tau_{x,DL}$ and $\tau_{x,FL}$, since the canted kagome plane makes the spin current no longer parallel to the $\mathbf{x}$-direction, which is consistent with the experimental results. In addition, the theoretical calculation of spin Hall conductivity for the $\Gamma_{4g}$ magnetic configuration performed in previous works also predicted the existence of $\mathbf{x}$-polarization~\cite{vzelezny2017spin, liu2019current}. Although the mechanism underlying the sign reversal behavior and the symmetry related to $\phi_{I}=45\si{\degree}$ for both torque ratios $\tau_{x,DL} / \tau_{y,DL}$ and $\tau_{x,FL} / \tau_{y,FL}$ in Fig.~\ref{fig:2}(e), (f) remain unknown, we presume that one possible factor is the contribution from the AFM domain configuration. In fact, an AFM domain configuration that modulates the SOT efficiency in the collinear AFM IrMn / Py bilayer was recently reported~\cite{zhou2019large}, which highlights the correlation between AFM domain configuration and SOT. It is worthwhile noting that the $\mathbf{x}$-polarization generated by spin-momentum locking also to be facet-dependent, as the triangular magnetic momentum points to a different crystalline orientation. These contributions from facet-dependent spin momentum locking and AFM domain configuration are extremely likely to give rise to the sign reversal behavior. Further research should be undertaken to investigate the correlation between non-collinear AFM domain configuration and the symmetry of the unconventional SOT to supplement what is reported above.

%% file: tex/03_conclusion.tex
In this paper, we have demonstrated the generation of an unconventional SOT (mainly $\mathbf{x}$-direction) in non-collinear AFM L1$_2$-ordered \mnpt{} with a topologically nontrivial spin structure. This unconventional $\mathbf{x}$-polarization in the SOT exhibited a unique symmetry with respect to the applied current direction, while both the unconventional SOT and the concomitant unique symmetry vanished in disordered \mnpt{}. Invariance of the sample with a Cu spacer excluded the possible effects from interlayer coupling, suggesting that the observation of unconventional SOT can be attributed to the MSHE and AFM domain configuration in L1$_2$-ordered \mnpt{}. Our findings provide the insight into how unconventional SOT can be generated by taking advantage of the topological spin structure in non-collinear AFM. In addition, this work offers new perspectives to clarify the underlying physics for charge to spin conversions in non-collinear AFMs.

%% file: tex/10_ackn.tex
This work was partially supported by the Center for Science and Innovation in Spintronics (CSIS), the Center for Spintronics Research Network (CSRN), and the GP-Spin program at Tohoku University. 